\documentclass[english,journal, draftcls, onecolumn]{IEEEtran}
\usepackage[T1]{fontenc}
\usepackage[latin9]{inputenc}
\usepackage{babel}
\usepackage{bm}
\usepackage{amsmath}
\usepackage{amssymb}
\usepackage{graphicx}
\usepackage[unicode=true,pdfusetitle,
 bookmarks=true,bookmarksnumbered=false,bookmarksopen=false,
 breaklinks=false,pdfborder={0 0 1},backref=false,colorlinks=false]
 {hyperref}
\begin{document}

\title{A Simple Method for 5G Positioning and Synchronization without Line-of-Sight}

\author{Henk Wymeersch\\
Department of Electrical Engineering\\
Chalmers University of Technology\\
41296, Gothenburg, Sweden\\
\texttt{henkw@chalmers.se}}
\maketitle
\begin{abstract}
In 5G mmWave, joint positioning and synchronization can be achieved
even when the line-of-sight (LOS) path is blocked. In this technical
note, we describe a simple method to determine a coarse estimate of
the user state and the environment. This method is based on geometric
consistency of the 5G mmWave measurements and can be used as a pre-processor
for other more sophisticated methods, in order to reduce their complexity.
A link to MATLAB source code is provided at the end of the document. 
\end{abstract}

\section{Introduction}

In 5G positioning, a received signal can carry information in the
delay and angle domain \cite{Shahmansoori2018}. A common model of
a received signal is 
\begin{equation}
\mathbf{y}(t)=\mathbf{W}^{\text{H}}\sum_{l=0}^{L}h_{l}\mathbf{a}_{\text{R}}(\bm{\theta}_{l})\mathbf{a}_{\text{T}}^{\text{H}}(\bm{\phi}_{l})\mathbf{s}(t-\tau_{l})+\mathbf{n}(t),\label{eq:ObservationModel1}
\end{equation}
where $\mathbf{s}(t)$ is a transmitted signal (possibly precoded),
$\mathbf{W}$ is a combining matrix, $h_{l}$ is a complex path gain,
$\bm{\theta}_{l}$ is the angle of arrival (AOA) in azimuth and elevation,
$\bm{\phi}_{l}$ is the angle of departure (AOD) in azimuth and elevation,
$\tau_{l}$ is the time of arrival (TOA), and $\mathbf{n}(t)$ is
(possibly colored) noise. The AOA and TOA are measured in the frame
of reference of the receiver, while the AOD is measured in the frame
of reference of the transmitter. The path index $l=0$ is the line-of-sight
(LOS) path, while the $L$ remaining paths are non-LOS (NLOS) paths. 

The AOA, TOA, and AOD of each path has a geometric meaning, which
depends on the location of the transmitter and receiver, as well as
the points of incidence of the NLOS paths in the environment (see
further). We will consider a downlink scenario, where the transmitting
base station (BS) has a known position, $\mathbf{x}_{\text{BS}}\in\mathbb{R}^{3}$,
the user equipment (UE) has an unknown position $\mathbf{x}\in\mathbb{R}^{3}$,
as well as an unknown clock bias $B$ (expressed in meters) and an
unknown orientation $\alpha$ with respect to the vertical axis. Moreover,
the unknown incidence points of the NLOS paths in the environment
are denoted by $\mathbf{x}_{s,l}\in\mathbb{R}^{3}$. These incidence
points or scattering points (SPs) could be related to reflecting surfaces
or small scattering objects. We further assume a channel estimation
routine is present at the receiver, which provides 
\begin{equation}
\mathbf{z}_{l}=[\tau_{l}\,\bm{\theta}_{l}^{\text{T}},\bm{\phi}_{l}^{\text{T}}]^{\text{T}}+\mathbf{w}_{l},\,\mathbf{w}_{l}\sim\mathcal{N}(\mathbf{0},\bm{\Sigma}_{l}).\label{eq:ObservationModel2}
\end{equation}

Our objective is now, given these measurements, to determine $\mathbf{x}$,
\textbf{$\alpha$}, $B$, $\{\mathbf{x}_{s,l}\}_{l=1}^{L}$ with only
\emph{minimal prior information} (the communication range $R$ between
transmitter and receiver) and in the \emph{absence of the LOS path}
(i.e., the path $l=0$ is not present, e.g., due to obstacles). 

\subsection{Related Works}

Similar problems were considered without $B$ in \cite{Shahmansoori2018},
where an exhaustive search over $\alpha$ was performed and for each
trial value a least-squares problems was solved to find an estimate
of $\mathbf{x}$ and $\{\mathbf{x}_{s,l}\}_{l=1}^{L}$. The estimate
with the lowest cost was then retained. In \cite{talvitie2017novel}
a Gibbs sampling approach was proposed, where first a sample of the
UE position and orientation was generated from a prior density, followed
by sampling the SPs, and so forth. In \cite{mendrzik2018joint}, a
factor graph method was proposed where particles are used to represent
the uncertainties of the unknown variables. For high measurement variance,
this method proved more robust than \cite{Shahmansoori2018}. While
there are many other works that relate to the problem at hand, only
those above consider the NLOS-only case. When the LOS is present the
problem becomes significantly more tractable (provided it is known
which path is the LOS path). When the LOS is absent and the transmitter
and receiver are not synchronized, the problem becomes significantly
harder: the lack of prior information leads to uninformative messages
in factor graph formulations, requiring a large number of particles
and high complexity. 

\subsection{Geometric Relations}

For completeness, the geometric relations between the location parameters
(BS, UE, and SPs) are provided:
\begin{itemize}
\item TOA: $\tau_{l}=\Vert\mathbf{x}_{s,l}-\mathbf{x}_{\mathrm{BS}}\Vert+\Vert\mathbf{x}_{s,l}-\mathbf{x}\Vert+B$
(times are normalized with the speed of light). 
\item AOA azimuth: $\theta_{l}^{(\text{az})}=\pi+\arctan2\left(y_{s,l}-y,x_{s,l}-x\right)-\alpha$,
where $\mathbf{x}_{s,l}=[x_{s,l},\,y_{s,l},\,z_{s,l}]^{\text{T}}$
and $\mathbf{x}=[x,\,y,\,z]^{\text{T}}$. 
\item AOA elevation: $\theta_{l}^{(\textrm{el})}=\arcsin\left((z_{s,l}-z)/\|\mathbf{x}_{s,l}-\mathbf{x}\|\right)$.
\item AOD azimuth: $\phi_{l}^{(\text{az})}=\arctan2\left(y_{s,l},x_{s,l}\right)$
\item AOD elevation: $\phi_{l}^{(\text{el})}$$=$$\arcsin\left((z_{s,l}-z_{\text{BS}})/\|\mathbf{x}_{s,l}-\mathbf{x}_{\text{BS}}\|\right)$,
where $\mathbf{x}_{\text{BS}}=[x_{\text{BS}},\,y_{\text{BS}},\,z_{\text{BS}}]^{\text{T}}$. 
\end{itemize}

\section{Proposed Method}

We start by making the following observations:
\begin{itemize}
\item The value of the AOD $\bm{\phi}_{l}$ and the UE bias $B$ define
a line segment from the BS to the $l$-th SP of length $\rho_{l}=\tau_{l}-B$.
This line is characterized by two endpoints: $\mathbf{x}_{\text{BS}}$
and $\mathbf{s}_{l}$, where 
\begin{equation}
\mathbf{s}_{l}=\mathbf{x}_{\text{BS}}+\rho_{l}\left[\begin{array}{c}
\cos\phi_{l}^{(\text{el})}\cos\phi_{l}^{(\text{az})}\\
\cos\phi_{l}^{(\text{el})}\sin\phi_{l}^{(\text{az})}\\
\sin\phi_{l}^{(\text{el})}
\end{array}\right].\label{eq:relation1}
\end{equation}
By construction, $\mathbf{x}_{s,l}$ is on this line segment. 
\item The values of the AOD $\bm{\phi}_{l}$ and the UE bias $B$ and the
UE orientation $\alpha$ define a line segment from the point $\mathbf{s}_{l}$
and a point $\tilde{\mathbf{s}}_{l}$, defined as 
\begin{equation}
\tilde{\mathbf{s}}_{l}=\mathbf{x}_{\text{BS}}+\rho_{l}\left[\begin{array}{c}
\cos\theta_{l}^{(\text{el})}\cos\left(\theta_{l}^{(\text{az})}+\alpha-\pi\right)\\
\cos\theta_{l}^{(\text{el})}\sin\left(\theta_{l}^{(\text{az})}+\alpha-\pi\right)\\
\sin\left(-\theta_{l}^{(\text{el})}\right)
\end{array}\right].\label{eq:relation2}
\end{equation}
By construction, $\mathbf{x}$ is on this line segment. 
\end{itemize}
Hence, $[\bm{\phi}_{l},B,\alpha]$ determine a line segment on which
the UE position must lie. Among these parameters, an estimate of $\bm{\phi}_{l}$
is directly available from (\ref{eq:ObservationModel2}). Since the
location of the UE, $\mathbf{x}$, must be consistent for all paths
$l=1,\ldots,L$, a unique intersection point among these $L$ lines
determines $\mathbf{x}$.

\begin{figure}
\includegraphics[width=0.32\columnwidth]{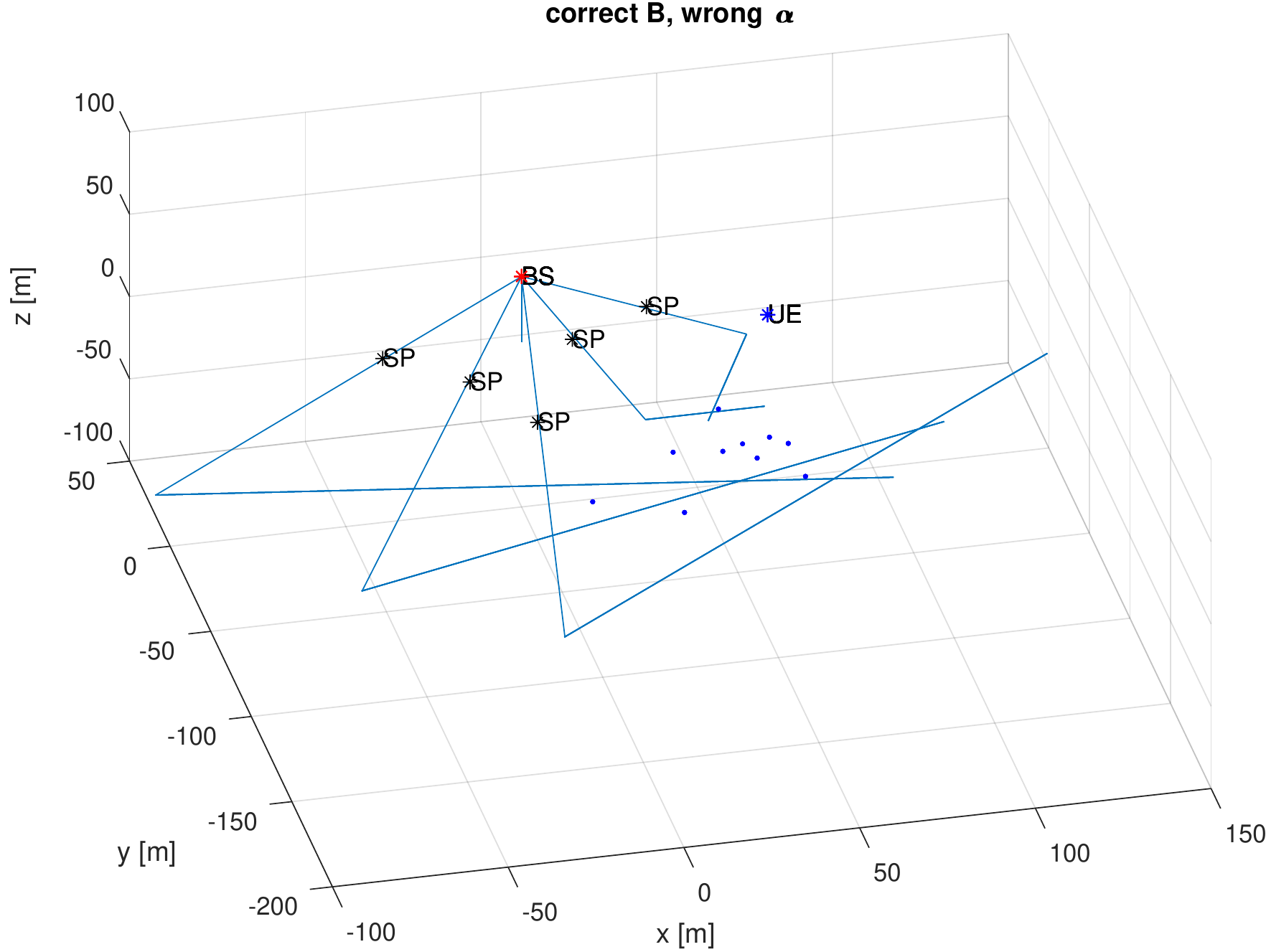}\hfill\includegraphics[width=0.32\columnwidth]{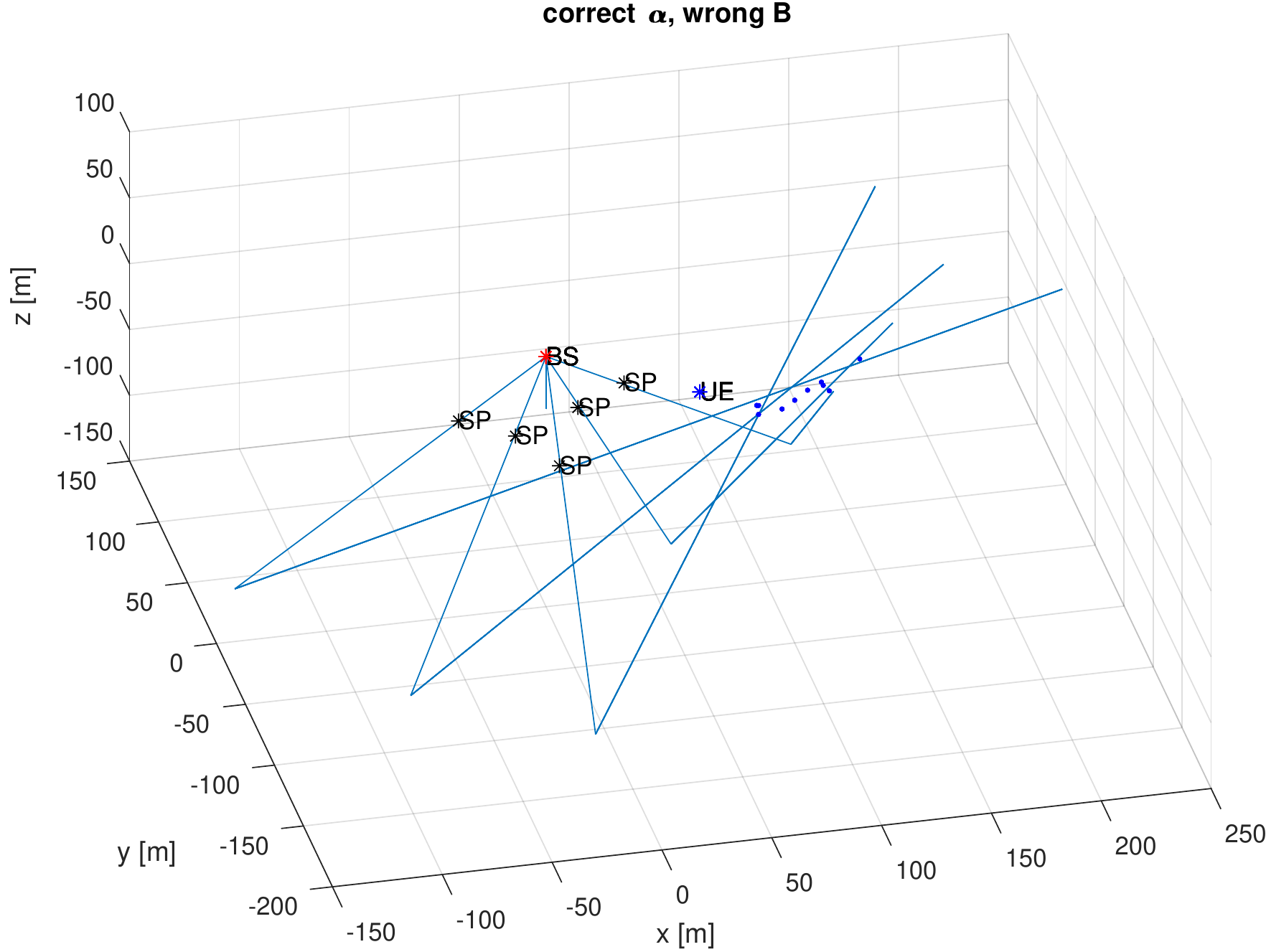}\hfill\includegraphics[width=0.32\columnwidth]{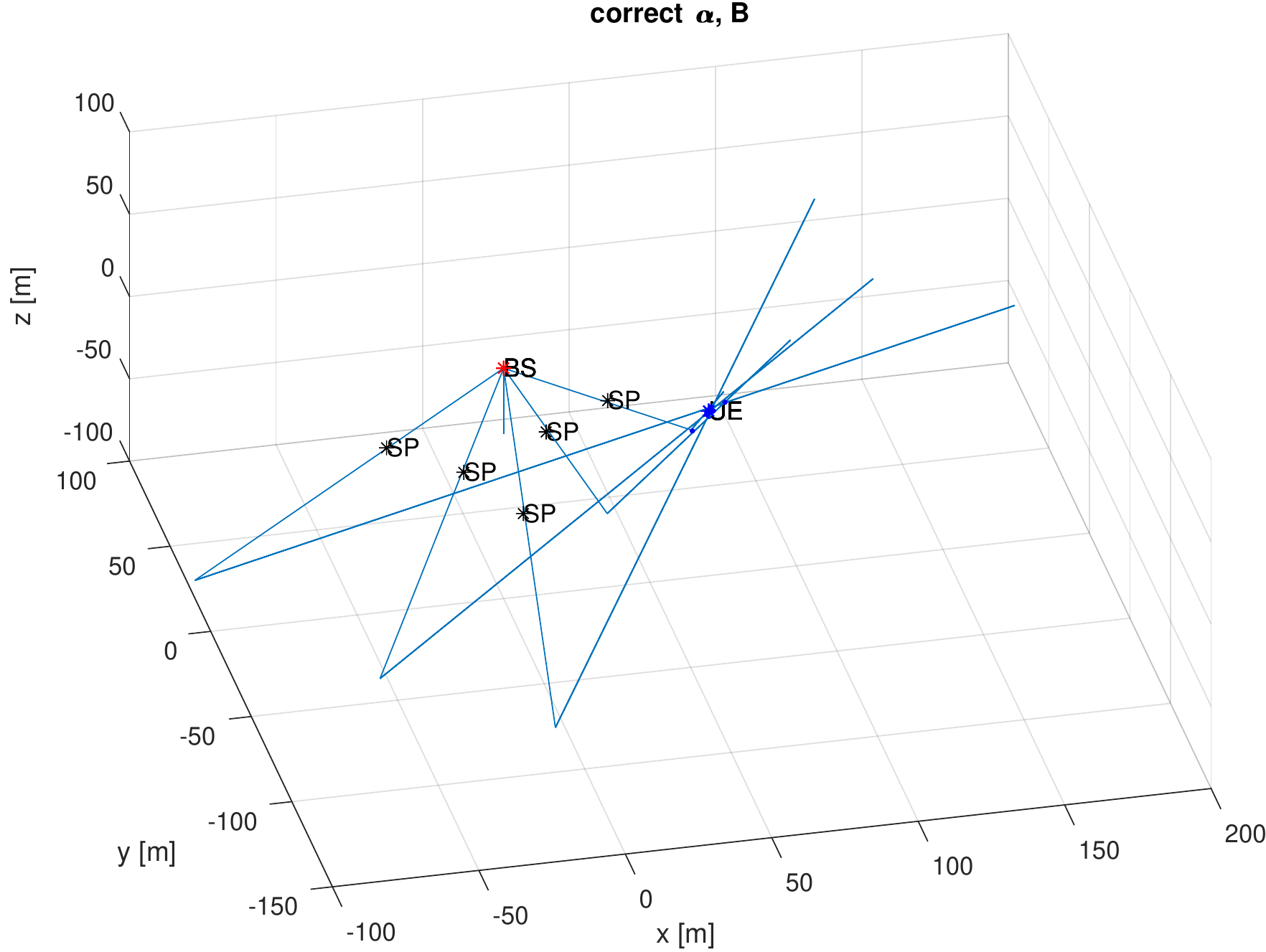}

\caption{\label{fig:Visualization-of-lines}Visualization of lines and intersection
points for different values of hypothesis ($\alpha,B$). The true
location of the UE, SPs, and the BS are shown, as are the samples
$\mathbf{x}_{l,l'}^{(n)}$ in blue dots. }

\end{figure}

While the above reasoning was performed in a noise-free case, the
noisy case can be treated by sampling:
\begin{enumerate}
\item Consider a hypothesis $(\alpha,B)$
\item For all $l=1,\ldots L$
\begin{enumerate}
\item Given $\mathbf{z}_{l}$, generate $N_{s}\ge1$ samples of $\bm{\phi}_{l}$
and $\rho_{l}$. For each sample $n$, generate a line segment as
described above, on which \textbf{$\mathbf{x}_{s,l}$} is hypothesized
to lie. 
\item Given $\mathbf{z}_{l}$, generate $N_{s}\ge1$ samples of $\bm{\phi}_{l}$,
$\rho_{l}$ and $\bm{\theta}_{l}$. For each sample $n$, generate
a line segment as described above, on which \textbf{$\mathbf{x}$}
is hypothesized to lie. We denote the segment by $\ell_{l}^{(n)}$
\end{enumerate}
\item For each sample $n=1,\ldots,N_{s}$, compute the distance $d_{l,l'}^{(n)}$
between $\ell_{l}^{(n)}$ and $\ell_{l'}^{(n)}$ , $l\neq l'$. At
the same time, determine the unique point in 3D at minimum distance
to $\ell_{l}^{(n)}$ and $\ell_{l'}^{(n)}$. Denote this point by
$\mathbf{x}_{l,l'}^{(n)}$. Fig.~\ref{fig:Visualization-of-lines}
shows these lines for a scenario with $L=5$ (leading to 10 lines)
and $N_{s}=1$. We observe that the points $\mathbf{x}_{l,l'}^{(n)}$
are more concentrated around the UE location when the hypothesis $(\alpha,B)$
is correct. 
\item Compute an error metric 
\[
\mathcal{E}(\alpha,B)=\frac{2}{N_{s}(L(L-1))}\sum_{n=1}^{N_{s}}\sum_{l=1}^{L}\sum_{l'>l}d_{l,l'}^{(n)}
\]
and a distribution of the UE location: 
\begin{align*}
\bm{\mu}_{\text{UE}}(\alpha,B) & =\frac{2}{N_{s}(L(L-1))}\sum_{n=1}^{N_{s}}\sum_{l=1}^{L}\sum_{l'>l}\mathbf{x}_{l,l'}^{(n)}\\
\bm{\Sigma}_{\text{UE}}(\alpha,B) & =\frac{2}{N_{s}(L(L-1))}\sum_{n=1}^{N_{s}}\sum_{l=1}^{L}\sum_{l'>l}(\mathbf{x}_{l,l'}^{(n)}-\bm{\mu}_{\text{UE}}(\alpha,B))(\mathbf{x}_{l,l'}^{(n)}-\bm{\mu}_{\text{UE}}(\alpha,B))^{\text{T}},
\end{align*}
where the expectation is computed using a sample average. 
\end{enumerate}
The best guess for $(\alpha,B)$ is then 
\[
(\alpha^{*},B^{*})=\arg\min_{(\alpha,B)}\mathcal{E}(\alpha,B),
\]
for which the UE position has been already computed. Finally, given
the UE position distribution and the value of $(\alpha^{*},B^{*})$,
it is straightforward to generate values of the SPs. 

\section{Numerical Results}

\begin{figure}
\centering

\includegraphics[width=0.5\columnwidth]{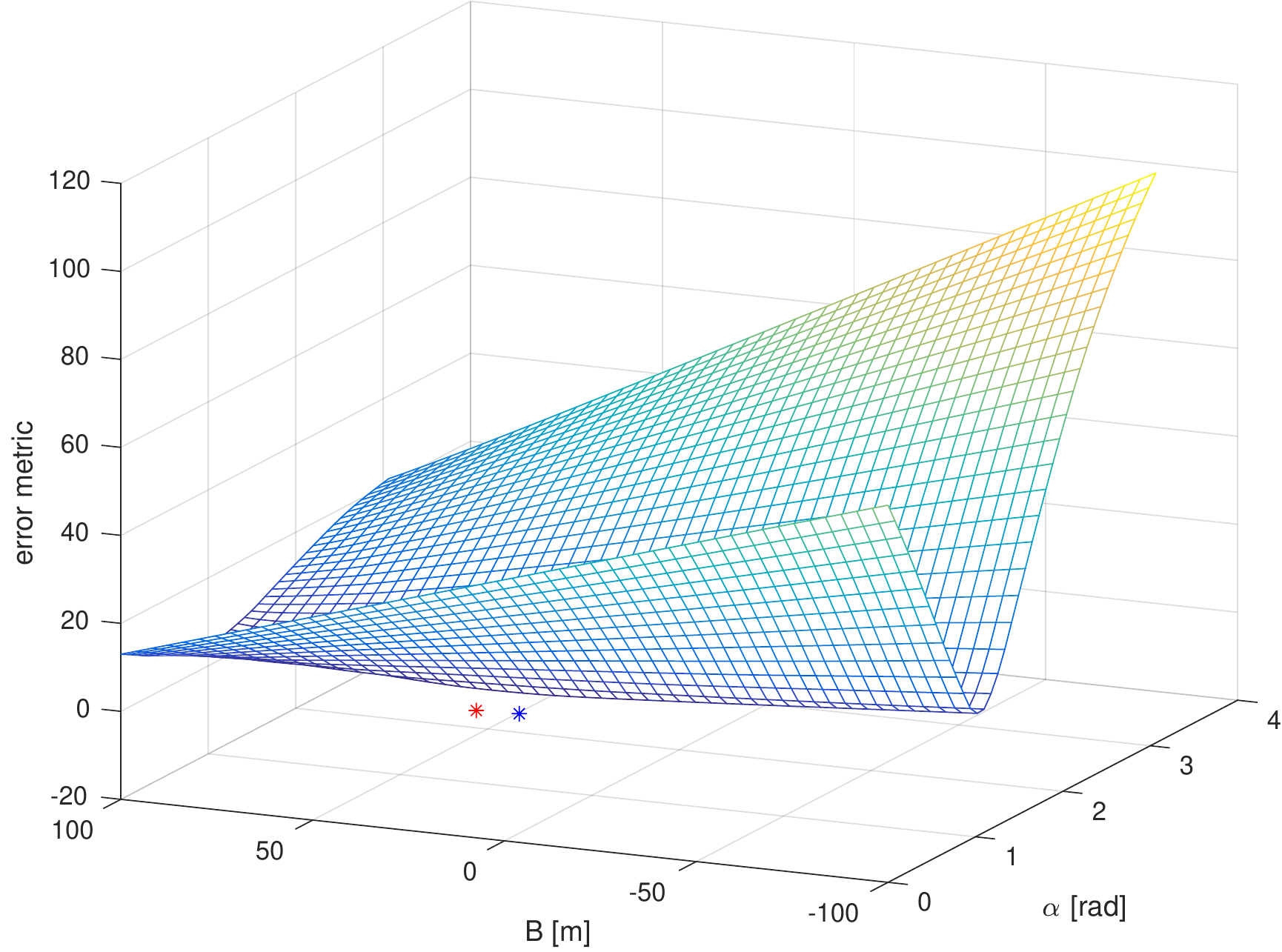}

\caption{\label{fig:Surface-plot-of}Surface plot of the error metric $\mathcal{E}(\alpha,B)$.
The true values were $\alpha=\pi/3$ and $B=20\,\text{m}$, marked
with a blue star. The estimated value $(\alpha^{*},B^{*})$ is marked
with a red star. }

\end{figure}
We consider a scenario with $L=5$ SPs and a measurement covariance
$\bm{\Sigma}_{l}=(\text{diag}[0.1,0.01,0.01,0.01,0.01])^{2},$i.e.,
10 cm TOA standard deviation and 0.01 rad AOD and AOA standard deviation.\footnote{Full details at \url{https://github.com/henkwymeersch/NLOSPosSynInit}.}
Using $N_{s}=10$ and a grid of values in the $(\alpha,B)$ space,
resulting error metric $\mathcal{E}(\alpha,B)$ is shown in Fig.~\ref{fig:Surface-plot-of}.
We make a number of observations. First of all, the error metric is
rather smooth and has a unique global minimum. The error function
is steep along the $\alpha$ axis but broad along the $B$ axis. This
is because a different value of $\alpha$ will lead to line segments
pointing in many different directions, leading to a larger error value.
In contrast, a different value of $B$ will lead to line segment that
nearly intersect, leading to only a very small increase in the error
metric. Overall, the figure indicates that a coarse search over $\alpha$for
an arbitrary value of $B$, followed by a gradient descent type method
can recover the global minimum. This indicates that it is easier to
determine the UE rotation than it is to synchronize the user. 

\begin{figure}
\centering

\includegraphics[width=0.5\columnwidth]{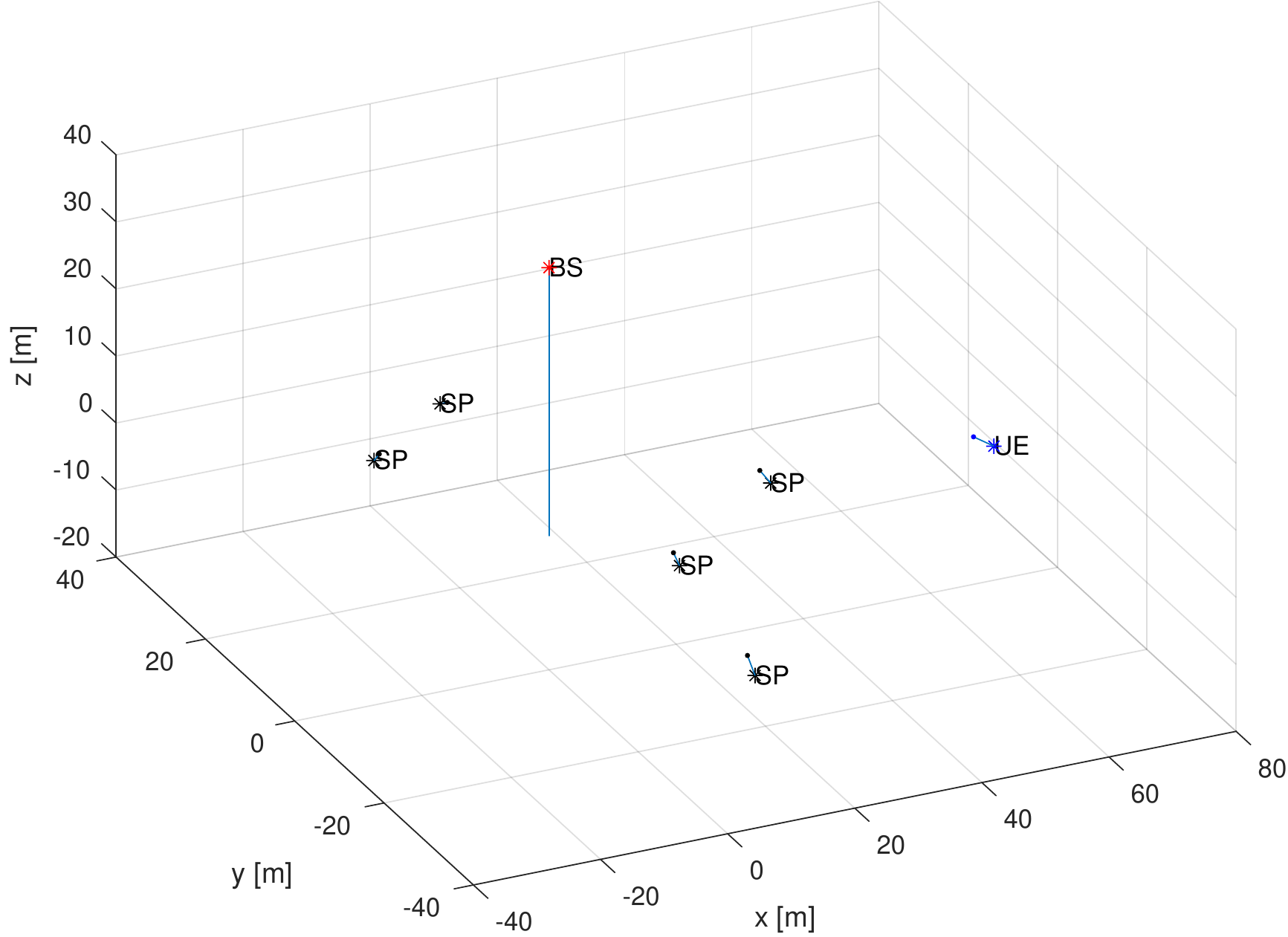}

\caption{\label{fig:Locations-of-the}Locations of the UE and the SPs, as well
as their estimates (connected with lines). }

\end{figure}
Next, after the global search is completed, we look into the quality
of the positioning and the map (determining the SP locations). The
result is shown in Fig.~\ref{fig:Locations-of-the}. We observe that
while we are able to determine all the unknowns, the error in the
clock bias leads to a shifting of the entire map towards the BS. Nevertheless,
we have been able to provide a rough initial estimate along with a
measure of uncertainty (not shown in the figure). 

\section{Conclusions}

We have presented a simple method to initialize downlink 5G positioning
in the challenging case where LOS is blocked, the position of the
user is unknown, there is no a priori map information, the user has
an unknown orientation, and the user and base station are not synchronized.
The method is based on a geometric consistency argument, where all
paths from the base station to the incidence points and to the user
must intersect in a fixed location (the user's location). Simulation
results indicate that it is much easier to determine the user's orientation
that it's clock bias. The method may have use as a pre-processor for
more refined positioning, synchronization, and mapping algorithms.
Complete source code is available at \url{https://github.com/henkwymeersch/NLOSPosSynInit}. 

\section*{Acknowledgment}

The author is grateful to discussions with Gonzalo Seco-Granados,
Nil Garcia, Hyowon Kim, and Rico Mendrzik. This work was supported,
in part, by the EU H2020 project 5GCAR, and the VINNOVA COPPLAR project,
funded under Strategic Vehicle Research and Innovation Grant No. 2015-04849. 

\bibliographystyle{ieeetr}
\bibliography{references}

\end{document}